\documentclass[twocolumn,english,aps,prl,showpacs,groupedaddress]{revtex4-1}
\usepackage{subfigure}
\usepackage{graphics}

\usepackage{graphicx}
\usepackage[T1]{fontenc}
\usepackage[latin9]{inputenc}
\usepackage{ifsym}
\usepackage{wasysym}
\usepackage{babel}
\usepackage{epstopdf}
\usepackage{color}

\begin{document}

\title{Mixing layer instability and vorticity amplification in a creeping viscoelastic flow}

\author{Atul Varshney$^{1,2}$ and Victor Steinberg$^{1,3}$}
\affiliation{$^1$Department of Physics of Complex Systems, Weizmann
Institute of Science, Rehovot 76100, Israel\\$^2$Institute of Science and Technology Austria, Am Campus 1, 3400 Klosterneuburg, Austria\\$^3$The Racah Institute of Physics, Hebrew University of Jerusalem, Jerusalem 91904, Israel}


\begin{abstract}
We report quantitative evidence of mixing-layer elastic instability in a viscoelastic fluid flow between two widely spaced obstacles hindering a channel flow at $Re\ll1$ and $Wi\gg1$.  Two mixing layers with nonuniform shear velocity profiles are formed in the region between the obstacles. The mixing-layer instability arises  in the vicinity of an inflection point on the shear velocity profile with a steep variation in the elastic stress. The instability results in an intermittent appearance  of small vortices in the mixing layers and an amplification of spatio-temporal averaged vorticity in the elastic turbulence regime. The latter is characterized through scaling of friction factor with $Wi$, and both  pressure and velocity spectra.   Furthermore, the observations reported provide  improved understanding  of the stability of the mixing layer in a viscoelastic fluid at large elasticity, i.e. $Wi\gg1$ and $Re\ll1$, and oppose the current view of  suppression of vorticity solely by polymer additives.
\end{abstract}

\maketitle

\section{Introduction}
Instability of a parallel shear flow of Newtonian fluid in different flow geometries and for various shear velocity profiles, particularly in connection to the transition to turbulence, was a subject of extensive investigation for about 150 years since first studies of Kelvin and Helmholtz \cite{batchelor,drazin,kundu}. A subclass of such flows, namely free shear flows, includes jets, wakes and mixing layers. The simplest of them is the mixing layer $-$ a layer between two uniform fluid streams with boundaries located far away $-$ which is distinguished by a nonlinear shear velocity profile with an inflection point  \cite{batchelor,drazin,kundu}.  At moderate Reynolds number $Re$, the mixing layer becomes unstable due to  Kelvin-Helmholtz (KH) instability resulting in a vortex chain $-$  classical example of the vortex generation in otherwise irrotational flow \cite{batchelor,drazin,kundu,saffman92,landau}.  A further increase of $Re$ results in vorticity amplification due to vortex stretching by a velocity gradient, the main mechanism of the vorticity enhancement in Newtonian fluid flow \cite{saffman92,landau}. Moreover, it is shown that the KH instability of the free shear layer plays a key role in a self-sustained process in  transition to turbulence in pipe, channel and plane Couette flows, where the original streamwise rolls supply energy to a streak flow that becomes unstable due to the KH instability resulting in  re-energizing the original streamwise rolls  \cite{waleffe}.

In viscoelastic fluids, the role of elastic stress, generated by polymers stretching in a shear flow, in the stability of the free shear layer was  investigated experimentally as well as theoretically mostly at large Reynolds $Re$ and Weissenberg numbers ($Wi$),  and at low polymer concentrations \cite{bird}. Here, $Wi$ defines a degree of polymer stretching. A general conclusion is drawn from  studies prior to 1990's  that no pure elastic instabilities exist in the parallel shear viscoelastic flows \cite{larson}. The first stability analysis of a viscoelastic fluid planar mixing layer, formed by two fluid streams moving with equal speeds in opposite directions,  shows that the elastic stress has a stabilizing effect for $Re\gg1$ and $Wi\gg1$ \cite{homsy}, in agreement with the results of early experimental studies \cite{hibberd,scharf}. In an insightful theoretical study, Hinch considered an elastic membrane between two counter-propagating Newtonian fluid streams, similar to two-fluid Newtonian streams with surface tension at the interface (see appendix in Ref. \cite{homsy}).  And it was found that the inertial instability of a Newtonian mixing layer is stabilized by a surge in elastic stresses across the interface for high elasticity ($El=Wi/Re$). In the elastic fluid, polymers stretched by a shear flow forms an elastic membrane that resists bending. For small non-zero elasticity, a new instability driven by a discontinuity in the normal elastic stress was reported \cite{rallison}.   Further, a suppression of the inertial instability by elasticity in the mixing layer of a polymer solution was confirmed by experiments on tunnel flow with a sudden large backward facing step, where a delay in the formation of  large-scale structures is observed \cite{cadot1}, as well as in wakes \cite{cadot2,cadot3,goldburg} at $Re\gg1$ and $Wi\gg1$. However, a  recent study extended to spatially developing mixing layers \cite{zaki} reveals, in contrast, the destabilizing effect of elastic stress at moderate values of $El$, while at large $El$ the flow is stabilized in accord with \cite{homsy}. Moreover,  these results ruled out the occurrence of a pure elastic instability in  free shear flows.

Here, we concern with a pure elastic instability of the mixing layer at $Re\ll1$ and $Wi\gg1$, i.e., an elastic analog of the Kelvin-Helmholtz instability in viscoelastic creeping flows. To the best of our knowledge,  theoretical as well as experimental studies are lacking in the literature on this subject. However, very recent numerical studies, still unpublished, show a generic  pure elastic instability of the mixing layer with a hyperbolic tangent shear velocity profile leading to a large jump in the elastic stress \cite{morozov,searle}, similar to the investigated in Ref. \cite{homsy} but conducted at $Re\ll1$ and $Wi\gg1$.  Moreover, at $Re\gg1$ the numerical results in Ref. \cite{morozov,searle} confirm the conclusion of Ref. \cite{homsy} that elasticity stabilizes the mixing layer, while at $Re\ll1$ and fixed $Wi\gg1$, the new pure elastic instability leads to a generation of a vortex chain state.  As indicated in \cite{searle}, the elastic KH instability could be  a generic mechanism in a self-sustained process in  transition to elastic turbulence in channel and plane Couette flows at $Wi\gg1$ and $Re\ll1$ \cite{morozov1}, similar to that found and analyzed for Newtonian fluid at $Re\gg1$ \cite{waleffe}.

In this paper, we report quantitative evidence of elastic instability of two mixing layers in a viscoelastic fluid between two widely spaced obstacles hindering the channel flow at $Re\ll1$ and $Wi\gg1$. The instability results in the generation of small size vortices and  vorticity enhancement by elastic stress in absence of inertia. The two mixing layers with nonuniform shear velocity profiles $u(y)$ are generated by a pair of counter-rotating elongated vortices that fill the space between two obstacles in length and width. The latter is resulted from the instability causing an increase of the vortex length with $Wi$ at a preserved streamline curvature \cite{atul}. We  characterize the circulation of the small vortices and its dependence on $Wi$, transition to elastic turbulence (ET) and ET properties \cite{groisman1,groisman2,groisman3} by measuring the friction factor, the average vorticity growth and frequency power spectra of both velocity and pressure fluctuations as a function of $Wi$.

\section{Experimental setup}
\begin{figure}[htbp]
\begin{center}
\includegraphics[width=8cm]{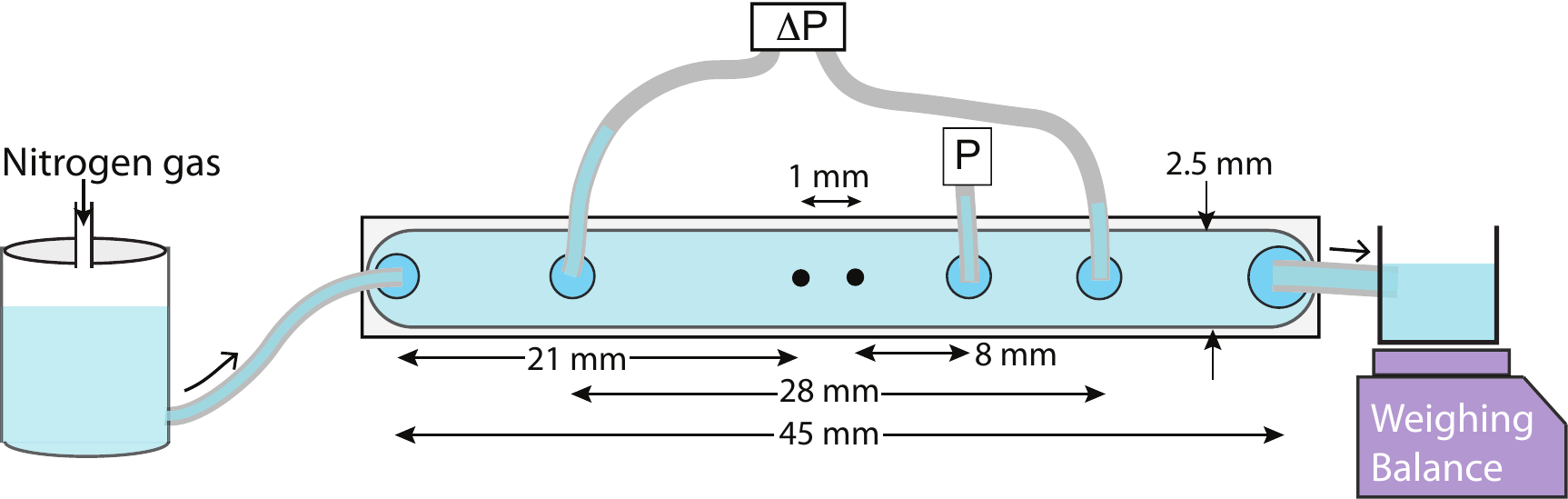}
\caption{Schematic of experimental setup (not to scale).}
\label{fig:setup}
\end{center}
\end{figure}

 The experiments are conducted in a linear channel of dimension $L\times w\times h=45\times2.5\times 1$  $mm^3$, shown schematically in Fig. \ref{fig:setup}. The fluid flow is hindered by two cylindrical obstacles of diameter $2R=0.30$ mm made of stainless steel separated by a distance $e=1$ mm and located at the center of the channel. Thus the geometrical parameters of the device are $2R/w=0.12$,  $h/w=0.4$ and $e/2R=3.3$ (see Fig. \ref{fig:setup}). The channel is prepared from transparent acrylic glass (PMMA).   The fluid is driven by $N_2$ gas at a pressure up to $\sim 10~psi$ and injected via the inlet into a rectangular channel. As a working fluid, a dilute polymer solution of high molecular weight polyacrylamide (PAAm, $M_w=18$ MDa; Polysciences) at concentration $c=80$ ppm ($c/c^*\simeq0.4$, where $c^*=200$ ppm is the overlap concentration for the polymer used \cite{liu2}), is prepared in a viscous solvent of $60\%$ sucrose and $1\%$ NaCl by weight. The solvent viscosity, $\eta_s$, at $20^{\circ}$C is measured in a commercial rheometer (AR-1000; TA Instruments) to be $0.1~Pa\cdot s$. An addition of the polymer to the solvent increases the solution viscosity, $\eta$, up to $0.13~Pa\cdot s$. The stress-relaxation method \cite{liu2} is employed to obtain longest relaxation time of the polymer solution, $\lambda=10\pm 0.5$ s.

 A differential pressure sensor (HSC series, Honeywell) measures the pressure drop $\Delta P$ across the obstacles and an additional absolute pressure sensor (ABP series, Honeywell) records the pressure $P$ fluctuations after the downstream cylinder, as shown schematically in Fig. \ref{fig:setup}. The fluid exiting the channel outlet is weighed instantaneously $W(t)$ as a function of time $t$ by a PC-interfaced balance (BA210S, Sartorius) with a sampling rate of $5$ Hz and a resolution of $0.1$ mg. The time-averaged fluid discharge rate $\bar{Q}$ is estimated as $\overline{\Delta W/\Delta t}$. Thus, Weissenberg and Reynolds numbers are  defined as $Wi=\lambda\bar{u}/2R$ and $Re=2R\bar {u}\rho/\eta$, respectively; here $\bar{u}=\bar{Q}/\rho wh$ and fluid density $\rho=1286~Kg/m^3$. For flow visualization, the solution is seeded with fluorescent particles of diameter $1~\mu m$ (Fluoresbrite YG, Polysciences). The region between the obstacles is imaged in the mid-plane {\it{via}} a microscope (Olympus IX70), illuminated uniformly with LED (Luxeon Rebel) at $447.5$ nm wavelength, and a CCD camera (GX1920; Prosilica) attached to the microscope records about 5000 images of a spatial resolution $1936\times 1456$ pixel at a rate of $50~fps$. To conduct particle image velocimetry ($\mu$PIV) with high spatially resolved velocity measurements, CMOS camera (MC124MG-SY with Sony sensor IMX253; XIMEA) is used  to record 600 frames with a spatial resolution of $4112\times3008$ pixel and 8 bits gray scale resolution  at a rate of $35~fps$ that allows us to obtain the spatially-resolved velocity field $\vec{U}=(u,v)$ of $5\times5~\mu m^2$ with $50\%$ overlap in the region between the cylinders for a chosen interrogation window $32\times 32~pixel^2$ \cite{piv}.
\begin{figure}[t!]
\begin{center}
\includegraphics[width=8.5cm]{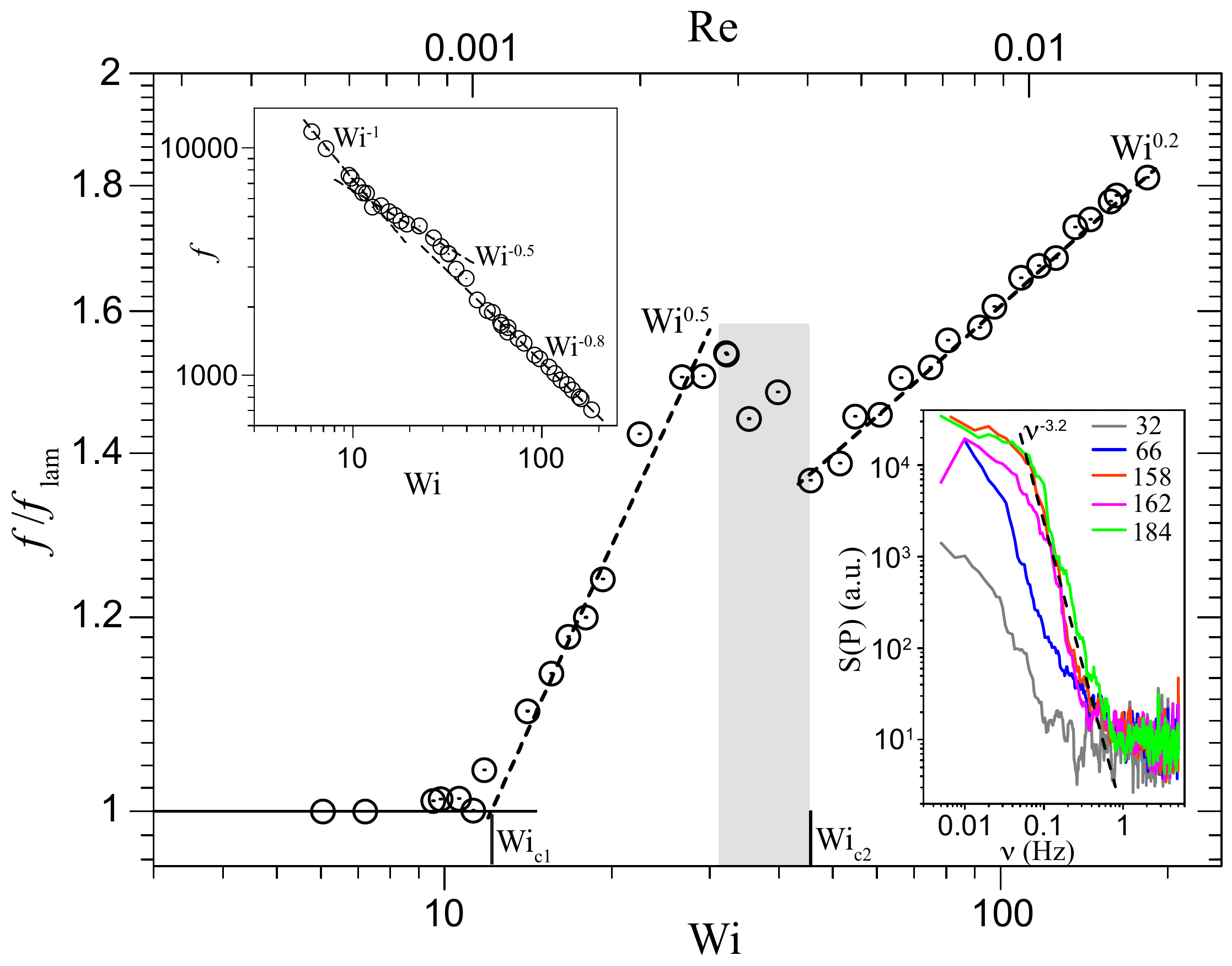}
\caption{Normalized friction factor $f/f_{lam}$ versus $Wi$ with the fits marked by dash lines above the first instability $fWi\sim Wi^{0.5}$ and in the ET regime $fWi\sim Wi^{0.2}$, which yield critical values for for respective transitions $Wi_{c1}\simeq12$  and $Wi_{c2}\approx 45$. Grey band indicates the transition region. Top inset: Friction factor $f$ versus $Wi$. Laminar flow is represented by scaling $f_{lam}\sim Wi^{-1}$. Bottom inset: Frequency power spectra of pressure fluctuations $S(P)$ in log-log coordinates for several $Wi$ values. Dashed line indicates the power spectra decay with an exponent $\beta\approx-3.2\pm0.2$  in the ET regime.}
\label{fig:dragcoeff}
\end{center}
\end{figure}

\section{Results}
\begin{figure}[htbp]
\begin{center}
\includegraphics[width=8.5cm]{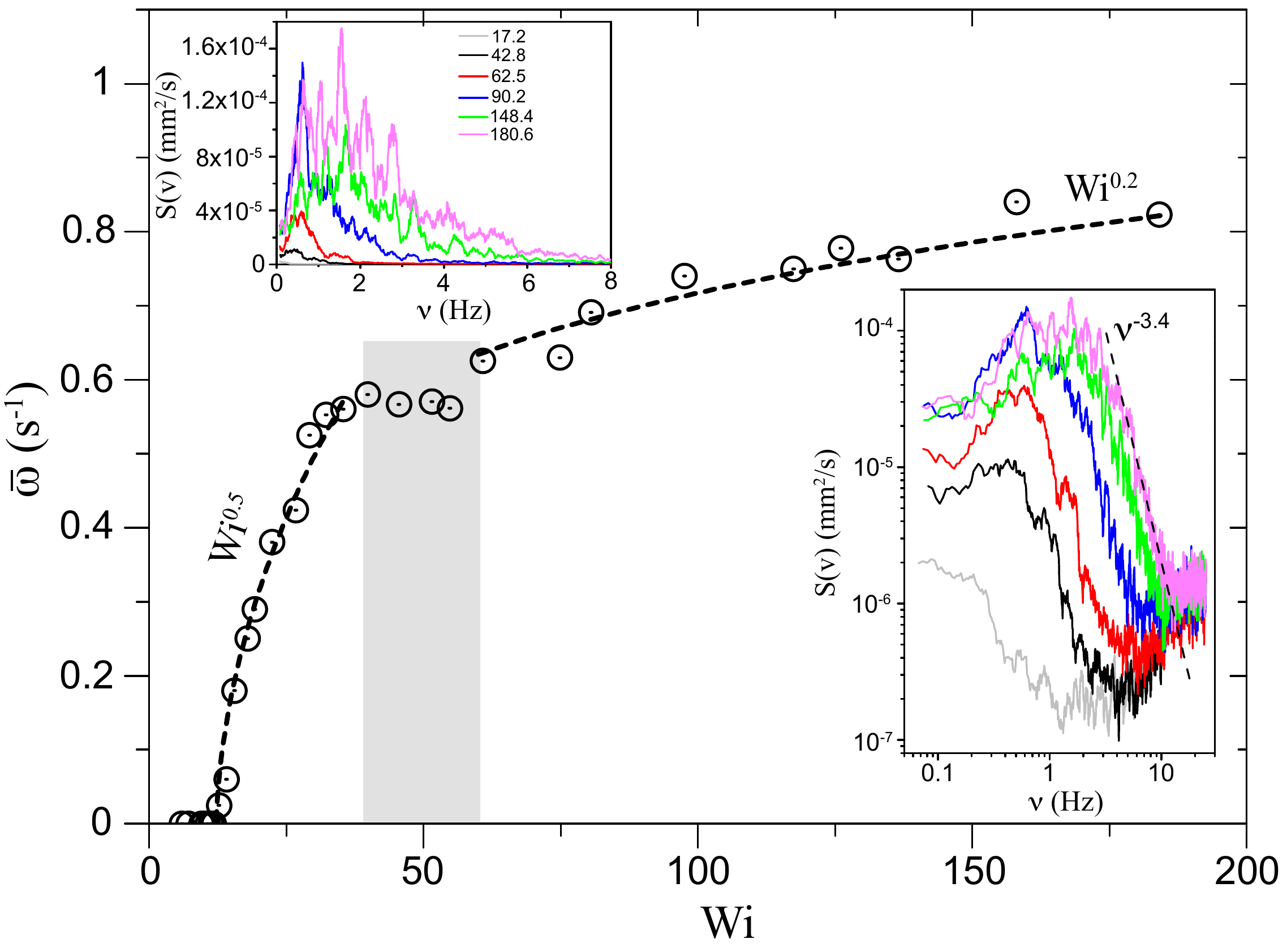}
\caption{ Spatially and time-averaged vorticity $\bar{\omega}$ versus $Wi$ with the fits by dashed lines $\bar{\omega}\sim Wi^{0.5}$, and $\bar{\omega}\sim Wi^{0.2}$. The grey band indicates the transition region of small vortex generation. Top inset: Frequency power spectra of cross-stream velocity, $S(v)$, for several $Wi$ shown in linear coordinates. Bottom inset:  $S(v)$ in log-log presentation for the same $Wi$. Dashed line indicates the power spectra decay with an exponent $\alpha\approx -3.4 \pm 0.1$ in the ET regime.}
\label{fig:avervorticity}
\end{center}
\end{figure}

 Various flow regimes are explored in the experiment up to maximum values of $Wi_{max}\simeq 184$ and $Re_{max}\simeq 0.02$. Inset in Fig. \ref{fig:dragcoeff} shows the dependence of the friction factor versus $Wi$, and the main plot in Fig. \ref{fig:dragcoeff} presents the normalized friction factor $f/f_{lam}$ as a function of $Wi$, where $f=2D_h\Delta P/\rho \bar{u}^2 L_c$ is the friction factor, $D_h=2wh/(w+h)=1.43$ mm is the hydraulic radius, and $L_c=28$ mm is the distance between the tubes to measure the pressure drop in the channel (see Fig. \ref{fig:setup}). Above the elastic instability at $Wi_{c1}\simeq12$ and up to $Wi\approx30$, $f/f_{lam}$ grows with power-law $Wi^{0.5}$ that has been discussed in Ref. \cite{atul,note}, and another scaling region commencing at $Wi_{c2}\approx 45$ is characterized by scaling $f/f_{lam}\sim Wi^{0.2}$. These two scaling regions are separated by a transition region between $\sim30\leq Wi\leq 45$ marked by the grey band, where $f/f_{lam}$ reduces. Figure \ref{fig:avervorticity} shows the $Wi$-dependence of the spatial (region between the obstacles) and time ($\sim$100 s) averaged   vorticity $\bar{\omega}$. At $Wi_{c1}<Wi<30$, the power-law increase of $\bar{\omega}$ as $Wi^{0.5}$ above the forward bifurcation occurs due to the growth of two elongated counter-rotating vortices, as reported in Ref. \cite{atul}. In the transition region at $30\leq Wi\leq Wi_{c2}=45$, $\bar{\omega}$ slightly reduces, and at $Wi>Wi_{c2}$ the $\bar{\omega}$ grows as $Wi^{0.2}$ (see Fig.  \ref{fig:avervorticity}). The frequency power spectra of the cross-stream velocity $v$, taken at the location $(x/R, y/R)=(4.53, 0.14)$, exhibits  power-law decay with an exponent $\alpha\approx-3.4\pm0.1$ (bottom inset in Fig. \ref{fig:avervorticity}) at higher frequencies $\nu$ and at $Wi>Wi_{c2}$ $-$ characteristic to ET \cite{groisman3}. The frequency power spectra of the pressure fluctuations $S(P)$ in the same range of $Wi$ show scaling behavior $S(P)\sim \nu^{\beta}$, where $\beta\approx-3.2\pm 0.2$ (bottom inset in Fig. \ref{fig:dragcoeff}), which is also typical to the ET flow \cite{jun,jun1,atul1}. Moreover, at lower frequencies wide pronounced  peaks in $S(v)$ are manifested  at higher $Wi$ that indicates either noisy oscillations or wave propagation accompanied ET (top inset in Fig. \ref{fig:avervorticity}).

\begin{figure*}[htbp]
\includegraphics[width=14cm]{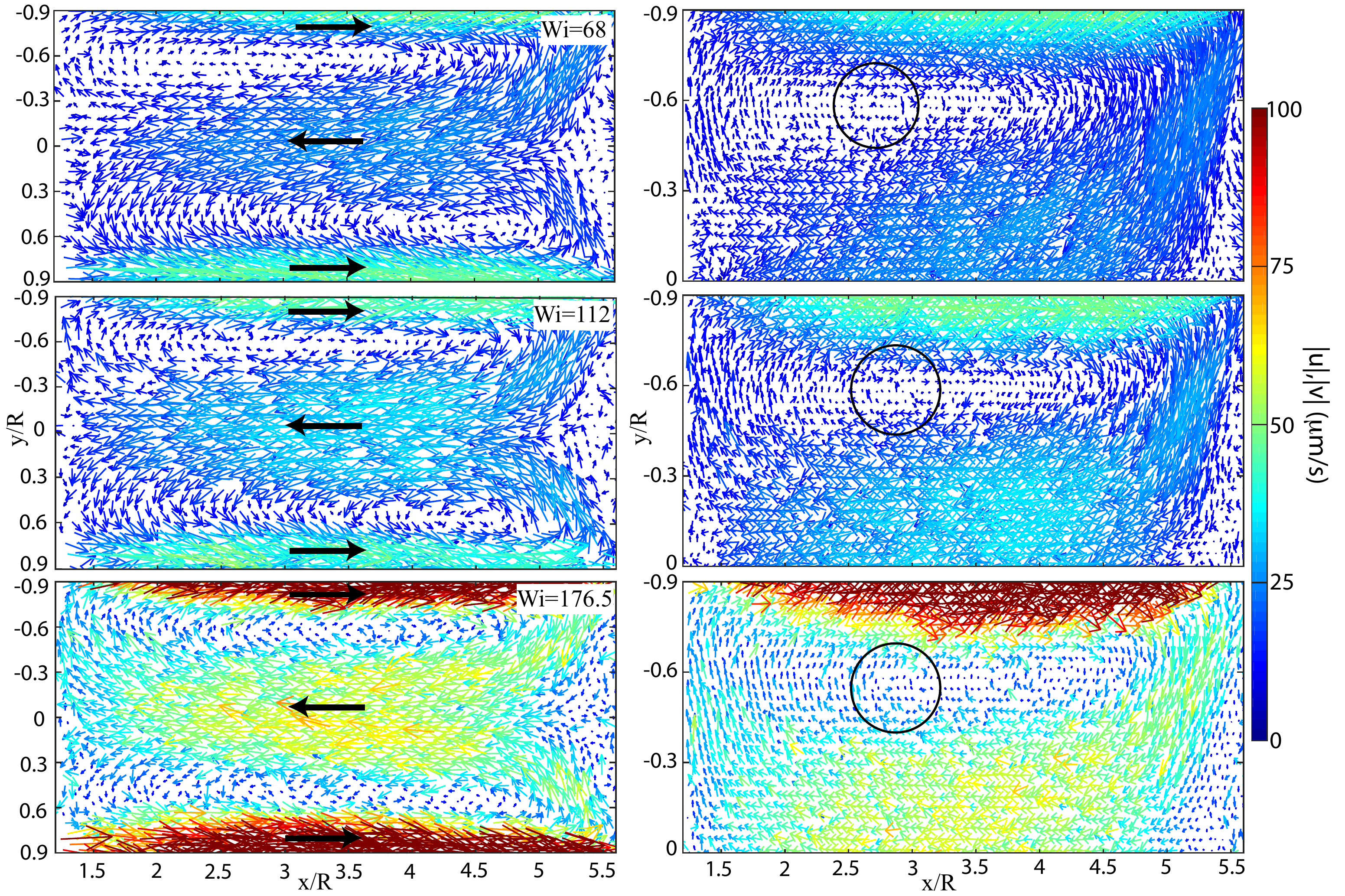}
\caption{Velocity fields between two obstacles at low (left panel) and high (right panel) magnifications for three values of $Wi$. The fields are obtained from $\mu$PIV and averaged over $\sim$1 s. Black arrows in left panel indicate the flow direction. Black circles in right panel highlight small vortex regions that are used  for circulation calculation.}
\label{fig:velfields}
\end{figure*}

To further illustrate the flow, snapshots of streak visualization averaged for $\sim$0.5 s are shown in Figs. 1SM, 2SM and 3SM at $Wi=40,~158$ and $184$, respectively, and also depicted through movies \cite{sm}, where small, about 40-60 micron size vortices can be clearly identified. First, one notices the presence of two - upper and lower - shear layers, where small vortices are located closer to a separatrix between inner and outer flow regions. Second, in the ET regime rare and strong bursts of perturbations originated at the inner surface of the downstream obstacle can be observed (see Movie 3 and a sequence of the snapshots in Fig. 3SM in \cite{sm}). A probable reason for these bursts may be overshooting of elastic stresses due to further stretching of polymers on curvilinear trajectories of the reverse flow.

Figure \ref{fig:velfields} shows velocity fields, time averaged $\sim$1 s, between two obstacles at low (left panel)  and high (right panel)  magnifications  for three values of $Wi$. In high magnification velocity field,  one can identify small vortices and one of the vortices for each $Wi$ is marked by a circle. One finds that vortices are located more on the right side, closer to the downstream obstacle. Due to  dynamical nature of the velocity field, intermittent strong bursts wash out the vortex structures and consequently the number of small vortices fluctuates  on the time scale of these bursts of $\tau_b\approx 1$ s. However, a long time averaging ($\sim$100 s) of velocity field shows a smooth two large vortex structure without any appearance of small vortices. Moreover, it is important to emphasize that the divergence of velocity field, i.e., $\nabla\cdot\vec{U}(x,y)$ fluctuates near zero in the whole region between the obstacles (see Fig. 4SM in \cite{sm} at $Wi=176.5$), which suggests the two-dimensional nature of the vortex flow.   
 
\begin{figure}[t!]
\begin{center}
\includegraphics[width=8.cm]{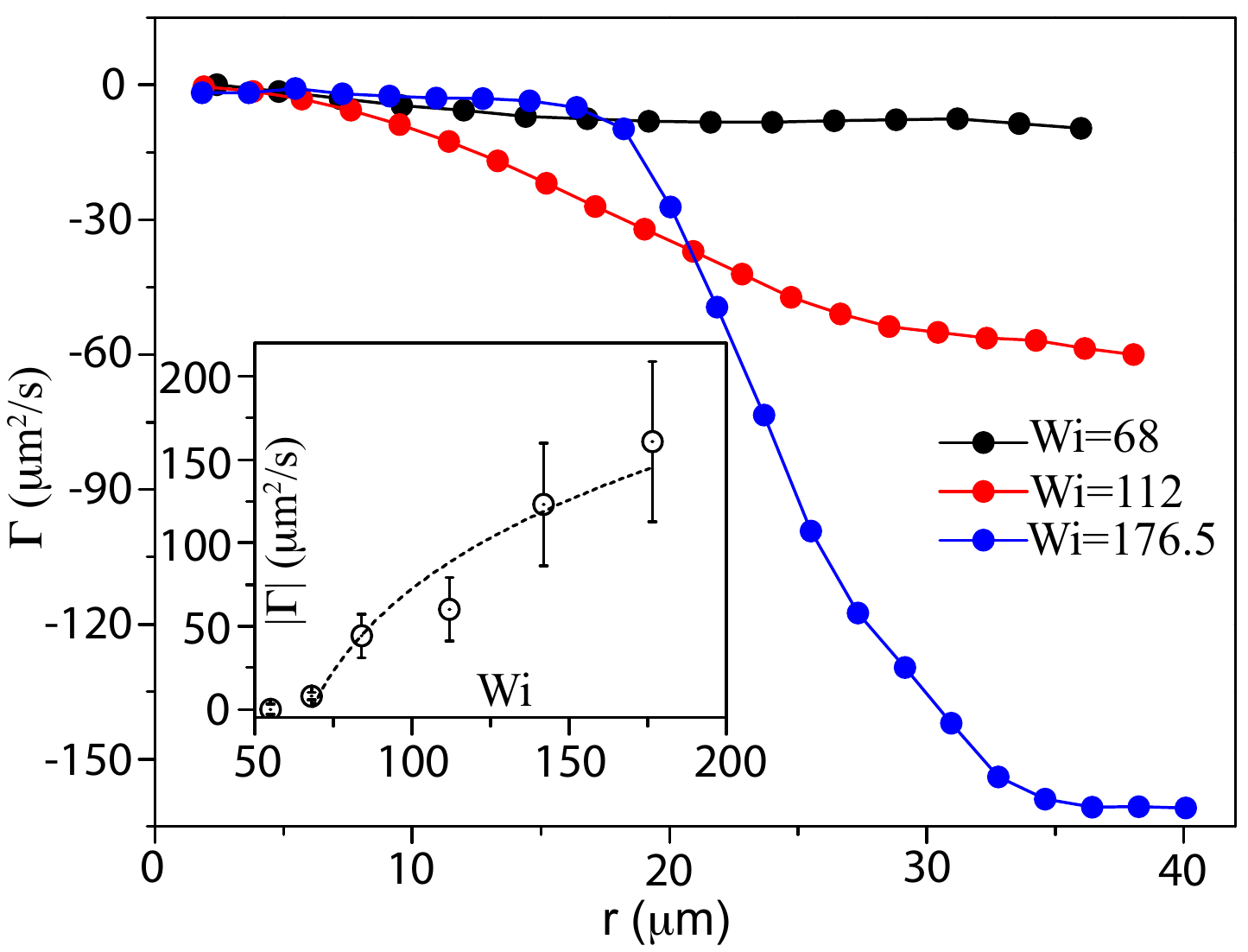}
\caption{Radial dependence of circulation $\Gamma$ for three values of $Wi$. Inset: variation of $|\Gamma|$ with $Wi$. The dashed line is a fit  $|\Gamma|\sim (Wi-55)^{0.2\pm0.05}$ to the data. }
\label{fig:circulation}
\end{center}
\end{figure}

 The small vortices are quantified through a calculation of the circulation $\Gamma=\oint_C{u_t ds}$, where $u_t$ is the azimuthal component of the velocity $\vec{U}$ integrated along a circle $C$. Fig. \ref{fig:circulation} shows the dependence of $\Gamma$ on the circle radius for three values of $Wi$. As  expected for an isolated vortex, $|\Gamma|$ grows with radius of a circle, on which the integration occurs, and then saturates at the radius value that depends on $Wi$. If an isolated vortex exists in uniform flow, $|\Gamma|$ first saturates and then reduces with further increasing  of integration radius, whereas for a vortex in a shear flow after the saturation $|\Gamma|$ should increase  due to the integration on the vorticity component of the shear. Indeed, this behavior is observed at further increase of the integration radius (see Fig. 5SM in \cite{sm}). Finally, in the inset of Fig. \ref{fig:circulation}, the saturated value of $|\Gamma|$ is plotted as a function of $Wi$ that indicates the onset of the small vortex generation at $Wi= 55\pm12$ ($\sim Wi_{c2}$) and shows above the onset a scaling behavior of the circulation (and so the vorticity) as $|\Gamma|\sim Wi^{0.2}$.

\begin{figure}[t!]
\begin{center}
\includegraphics[width=8.6cm]{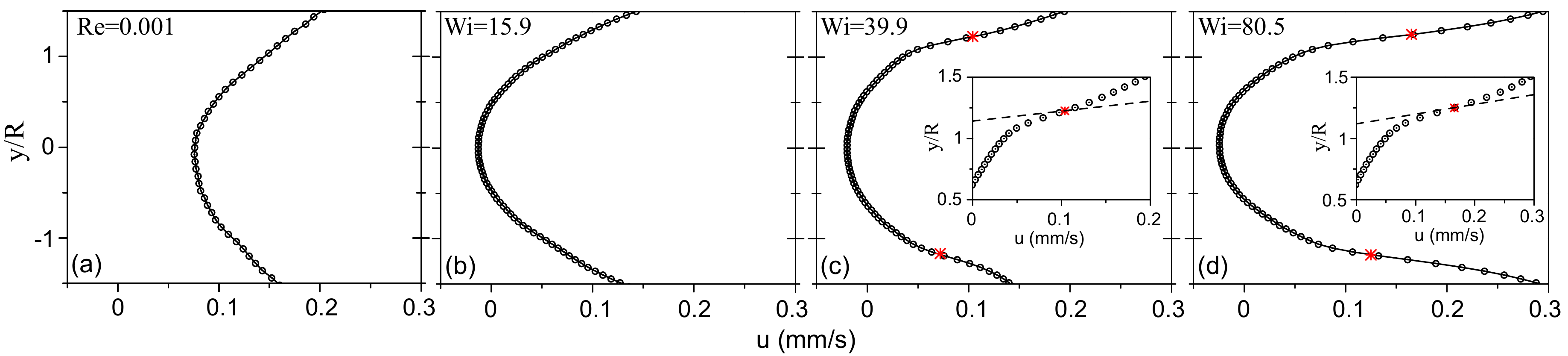}
\caption{Streamwise velocity profile $u(y)$ for (a) Newtonian solvent and (b-d) polymer solution at three $Wi$ values. Red symbols in (c-d) indicate the inflection points on the profiles. Insets in (c-d) show the positions of  inflection points  in details; dashed line is a tangent at the inflection point. Noticeably, the velocity profile just above the first instability at $Wi=15.9$ is similar to that in the laminar region of a Newtonian solvent.}
\label{fig:velprofile}
\end{center}
\end{figure}

\section{Discussion}
The main message delivered by the data is the onset of the generation of small vortices inside  two mixing layers at $Wi\approx Wi_{c2}$ in the ET regime and their number  fluctuates on the time scale $\tau_b$ of the occurrence of strong bursts. The circulation (and so the vorticity) of small individual vortices increases as $\sim(Wi-55)^{0.2}$ (inset in Fig. \ref{fig:circulation}), similar to the growth of the spatio-temporal averaged vorticity $\bar{\omega}$ of the large vortex  in the ET regime (see Fig. \ref{fig:avervorticity}). Thus, one observes three different regimes of $\bar{\omega}$ (Fig. \ref{fig:avervorticity}) in the wake of a viscoelastic creeping flow between two widely spaced obstacles hindering the channel flow: (i) growth of $\bar{\omega}$ as $Wi^{0.5}$ due to the increase of the vortex length at $Wi>Wi_{c1}$ above the elastic instability \cite{atul}, (ii) slight reduction of $\bar{\omega}$ in the transition region, and (iii) growth of $\bar{\omega}$ as $Wi^{0.2}$ in ET. The latter maybe associated with the growth of the circulation (and so the vorticity) of each individual small vortex with the similar scaling.

Similar to the variation of $\bar{\omega}$ with $Wi$, one finds three different regimes of the dependence of $f/f_{lam}$ on $Wi$ (see Fig. \ref{fig:dragcoeff}). A growth of $f/f_{lam}$ as $Wi^{0.5}$ at $Wi>Wi_{c1}$ above the elastic instability \cite{atul}. In the transition region $f/f_{lam}$ reduces with $Wi$, whereas the ET regime  is characterized by the power-law dependence $f/f_{lam}\sim Wi^{0.2}$. It may suggest that the $\bar{\omega}$ growth in the ET regime is the main cause for the increase of $f/f_{lam}$. Moreover, the variations of $\bar{\omega}$ and $f/f_{lam}$ are correlated in the whole range of $Wi$ from $Wi_{c1}$ and up to $Wi_{max}$ reached in the current study. An additional feature is the pronounced peaks at low frequencies in the velocity power spectra $S(v)$  indicating the presence of either oscillatory or wavy modes interacting with the vortices (see top inset in Fig. \ref{fig:avervorticity}). On the other hand, the scaling exponents of the velocity $\alpha\simeq -3.4$ and the pressure fluctuations $\beta\simeq-3.2$ power spectra at high $\nu$ provide a firm evidence of the ET regime \cite{groisman1,groisman2,groisman3,lebedev}.
\begin{figure}[htbp]
\begin{center}
\includegraphics[width=6cm]{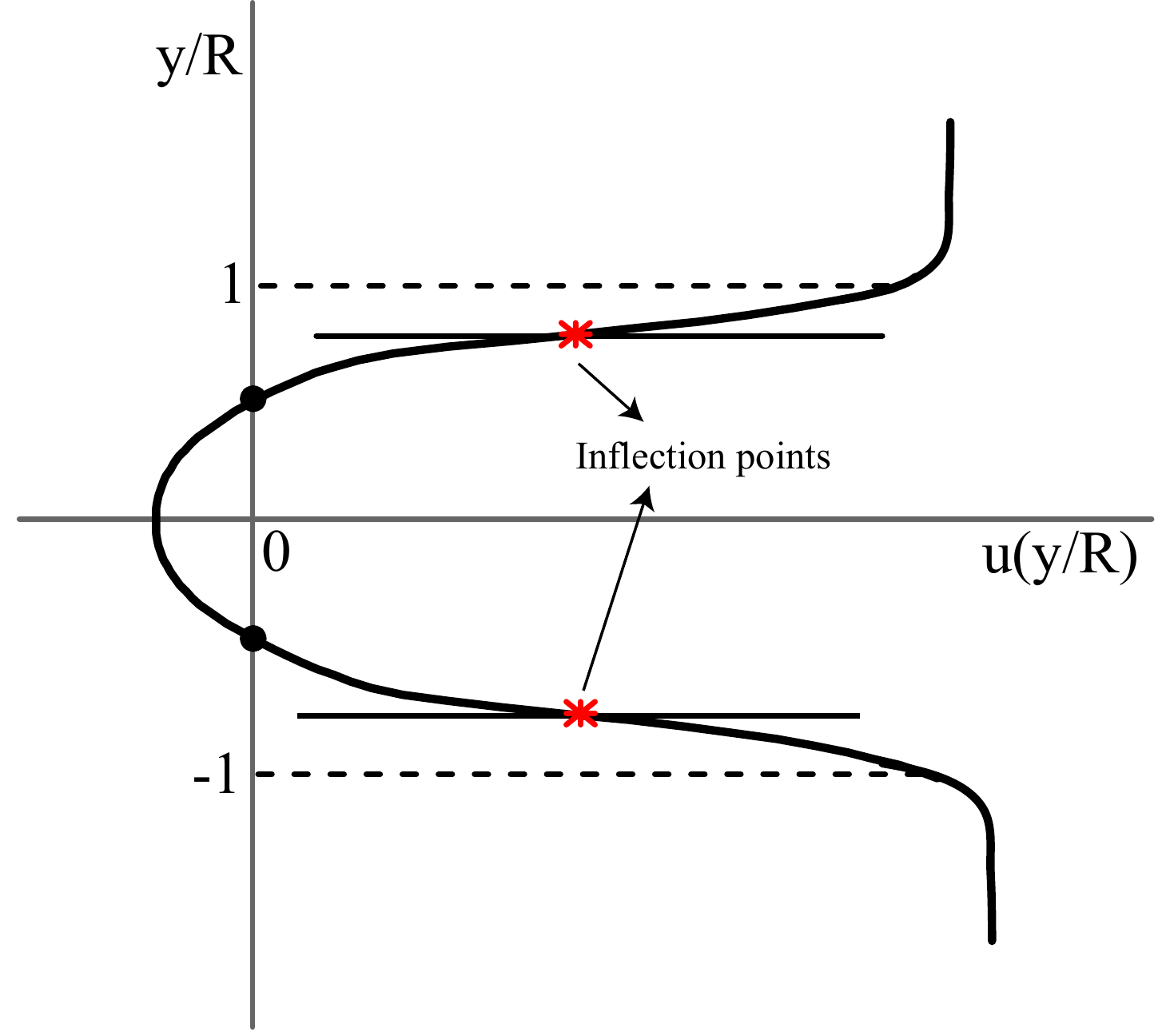}
\caption{Illustration of full streamwise velocity profile $u(y/R)$ in the inner and outer regions of the obstacles to demonstrate a similarity with two mixing layers. Red symbols are inflection points on the profile and solid lines are tangents at the inflection points. }
\label{fig:sketch}
\end{center}
\end{figure}

A natural question arises what can be the reason for the generation of the small vortices at $Wi>Wi_{c2}$ in the ET regime, what is a possible vorticity amplification mechanism in the ET regime and what is so unique in this flow configuration compared to other flow geometries, where ET was observed and investigated but the vorticity generation was not found \cite{groisman1,groisman2,groisman3,teo2,jun2}. In a mixing layer of a Newtonian fluid the presence of the inflection point in a streamwise shear velocity profile is the necessary condition for KH instability, called the Rayleigh criterion, and the inertia-driven KH instability of a mixing layer resulted in the generation of the vortex chain was theoretically understood and  observed in experiments as well \cite{batchelor,drazin,kundu}. In the case of  pure elastic instability, the presence of the inflection point in the nonuniform shear velocity profile suggests a steep variation in the elastic stress that may engender elastic instability resulting in the vortices generation.  Indeed at $Wi<\sim40$ a streamwise velocity $u(y/R)$ has a parabolic profile, whereas at $Wi>\sim40$  two symmetrical inflection points on $u(y/R)$ for each mixing layer are observed (see Fig. \ref{fig:velprofile} and also sketch in Fig. \ref{fig:sketch}). Thus the key question is whether  a mixing layer of a viscoelastic fluid at $Re\ll 1$  and $Wi\gg1$  with a nonuniform shear velocity profile containing an inflection point can be elastically unstable that results in the vortex generation. Perhaps it could be addressed through numerical simulations performed in this parameter regime and for the similar flow configuration.

\section{Conclusion}
Two key findings on the elastic instability of the mixing layer of a viscoelastic fluid at   $Re\ll1$ and $Wi\gg1$ are described. The observations are reminiscent of the Kelvin-Helmholtz instability of the mixing layer in a Newtonian fluid at $Re\gg1$ \cite{batchelor,drazin,kundu}, namely: (i) the vortex generation in two mixing layers in a viscoelastic  fluid between two obstacles hindering the channel flow at $Re\ll1$ and $Wi\gg1$, and (ii) the amplification of spatio-temporal averaged vorticity $\bar{\omega}$ and increase of the circulation $|\Gamma|$ of individual small vortices due to elastic stresses in the ET regime. Both observations provide a better understanding of the stability of  shear flow  of viscoelastic fluid at  $Wi\gg1$, $Re\ll1$ and $El\gg1$ \cite{larson,homsy,rallison} and in contrast with a well-known results that an injection of polymer additives into a Newtonian fluid flow inhibits vorticity \cite{cadot2,cadot3,goldburg} at $Wi\gg1$, $Re\gg1$.  Recent numerical studies, though unpublished,  conducted at $Wi\gg1$ and $Re\ll1$ reveal a generic  purely elastic instability in a mixing layer of a viscoelastic fluid with a hyperbolic tangent shear velocity profile leading to a large jump in the elastic stress and a generation of a vortex chain \cite{morozov,searle}, similar to the KH instability in a mixing layer of a Newtonian fluid.  In regard to the second finding, the mechanism of the vorticity enhancement by an elastic stress was suggested in Ref. \cite{zaki1} and maybe  relevant  to explain the second paradox. The recent theoretical and numerical studies reveal that the elastic stress is also capable to amplify a spanwise vorticity in a homogeneous viscoelastic shear flow particularly at $El\gg1$. Moreover, the polymer torque resulting from the elastic stress is amplified as vorticity perturbations become aligned with the shear \cite{zaki1}.

\section{Acknowledgements}
V. S. is grateful for illuminating discussions with V. Lebedev and A. Morozov. We thank Guy Han and Yuri Burnishev for technical support.  A.V. acknowledges support from the European Union's Horizon 2020 research and innovation programme under the Marie Sk{\l}odowska-Curie grant agreement No. 754411. This work was partially supported by the Israel Science Foundation (ISF; grant \#882/15) and the Binational USA-Israel Foundation (BSF; grant \#2016145).


\begin{thebibliography}{40}
\bibitem{note}{A lower value of $Wi_{c1}$ found in this experiment compared with that reported in \cite{atul} can be explained by a different base for the $\Delta P$ measurements: in \cite{atul} $\Delta P$ was measured between the inlet with a narrow metal capillary inside the inlet tube and the tube location downstream from the second obstacle, whereas in the current experiment it is measured across the obstacles in the channel. }
\end{thebibliography}
\end{document}